\documentclass[aps,prb,twocolumn,amsmath,amssymb,nofootinbib,eqsecnum,superscriptaddress,floatfix]{revtex4}

\usepackage{amsmath}
\usepackage{amssymb}
\usepackage[dvips]{color}
\usepackage{graphicx}
\usepackage{dcolumn} 
\usepackage{bm} 
\usepackage[hypertex]{hyperref}
\usepackage{longtable}

\begin{document}


\title{
Deconfined fractional electric charges in graphene at high magnetic fields
      }

\author{Chang-Yu Hou}
\author{Claudio Chamon}
\affiliation{
Physics Department, Boston University, Boston, MA 02215, USA
        }

\author{Christopher Mudry} 
\affiliation{
Condensed Matter Theory Group,
Paul Scherrer Institut, CH-5232 Villigen PSI, Switzerland
        }

\date{\today}

\begin{abstract}
The resistance at the charge neutral (Dirac) point was shown by
Checkelsky \textit{et al} in Phys.\ Rev.\ B \textbf{79}, 115434 (2009)
to diverge upon the application of a strong magnetic field normal to
graphene. We argue that this divergence is the signature for a
Kekul\'e instability of graphene, which is induced by the magnetic
field. We show that the strong magnetic field does not remove the zero
modes that bind a fraction of the electron around vortices in the
Kekul\'e dimerization pattern, 
and that quenched disorder present in
the system makes it energetically possible to separate the fractional
charges. These findings, altogether, indicate that graphene can
sustain deconfined fractionalized electrons.
\end{abstract}

\maketitle

\section{Introduction}

The elementary excitations in the fractional quantum Hall (FQH) effect
carry a fraction of the charge of the electron, as argued by
Laughlin.~\cite{Laughlin83} Experimental evidence for the fractionally
charged quasiparticles have been presented for the $\nu=1/3$ state,
using shot noise measurements in Refs.~\onlinecite{Glattli} and
\onlinecite{Reznikov} and by using a scanning tunneling transistor in
Ref.~\onlinecite{Yacoby}. It has remained an open question whether
there are experimental systems in two-dimensions (2D) other than the
FQH states for which the elementary excitations carry a fractional
charge~\cite{Rokhsar88,Moessner01} and for which the fundamental mechanism for electron
fractionalization is different from that in the FQH effect.

Recently, proposals for a mechanism to fractionalize the electron have
been suggested for graphene-like systems,~\cite{Hou07} in which the
electrons disperse according to the Dirac equation in 2D. The
mechanism involves opening a mass gap in the Dirac equation, via the
spontaneous breaking of a symmetry, 
the formation of a Kekul\'e bond
dimerization pattern. This mechanism is 
fundamentally different than the
one in the FQH effect. 
Instead, it is closer conceptually to the mechanism 
for fractionalization in 1D,~\cite{Jackiw76,Su79} 
which for 30 years was believed to be peculiar to 1D and not possible in 2D.
However, the strengths of the interactions in graphene are, alone,
insufficient to lead to the Kekul\'e instability. Moreover, even if the
Kekul\'e dimerization pattern formed,  
an axial gauge potential is needed
to deconfine the fractionally
charged quasiparticles that attach to vortices in the 
Kekul\'e pattern.~\cite{Jackiw07}

We are going to show in this paper that the presence of a magnetic
field in graphene stabilizes a quasi-long-range order in the form of a
Kekul\'e distortion associated to a U(1) continuous
symmetry,~\cite{Hou07,Chamon00} and opens an electronic energy gap
$2\Delta^{\ }_{0}$. This quasi-long-range U(1) order can be destroyed
by the unbinding of vortices due to either thermal- or
disorder-induced fluctuations. We then show that
fractional charges do attach to vortices in the Kekul\'e pattern even
in the presence of a strong magnetic field. Quenched disorder present
in the system, remarkably, makes it energetically possible to separate
the fractional charges. These theoretical results
on the formation of the Kekul\'e pattern in the
presence of a magnetic field and of static disorder explain the
features observed in the experimental measurements by Checkelsky
\textit{et al}.~\cite{Checkelsky08,Checkelsky09} 

Let us briefly discuss the known phenomenology of graphene in the
presence of a magnetic field. Starting from the linearized Dirac
spectrum 
$\varepsilon(\boldsymbol{k})= \pm\hbar v^{\ }_{F}|\boldsymbol{k}|$
around the two non-equivalent Fermi points (valleys)
$\boldsymbol{K}^{\ }_{+}=-\boldsymbol{K}^{\ }_{-}$,~\cite{Wallace47}
the single-particle Landau levels in a uniform applied magnetic field
$B$ perpendicular to graphene are
\begin{equation}
\varepsilon^{\ }_{n}=
\mathrm{sgn}(n)\,\hbar\omega^{\ }_{\mathrm{c}}\,\sqrt{|n|}.
\label{eq:4-fold-deg-Landau-levels}
\end{equation}
Here, $n$ is an integer, the cyclotron frequency is
$\omega^{\ }_{\mathrm{c}}=\sqrt{2}v^{\ }_{F}/\ell^{\ }_{B}$, while the
magnetic length is $\ell^{\ }_{B}=\sqrt{\hbar c/(eB)}$.%
~\cite{McClure56} This single-particle spectrum 
leads to the quantization of the dimensionless Hall conductivity
\begin{equation}
\nu\equiv
h\sigma^{\ }_{xy}/e^{2}=
4\times(n+1/2)=\pm2,\pm6,\pm10,\cdots,
\label{eq:even-quantization-sigmaxy}
\end{equation}
if the Zeeman splitting and the Coulomb or electron-phonon
interactions are neglected,%
~\cite{Deser82,Niemi83,Redlich84,Ishikawa84,Gusynin05} 
as observed experimentally for magnetic fields of up to
8 [T].~\cite{Novoselov05,Zhang05} 
On the other hand, the observation in Refs.~\onlinecite{Zhang06}
and \onlinecite{Jiang07} of new plateaus at the filling fractions
$\nu=0,\pm 1,\pm 4$, for applied magnetic fields between 
20 [T] and 40 [T]
indicates that the 4-fold degeneracies of the Landau
levels~(\ref{eq:4-fold-deg-Landau-levels}) must be lifted. For
example, spin-splitting may be induced by the Zeeman or Coulomb
interactions while valley-splitting ($\boldsymbol{K}^{\ }_{+}$ and
$\boldsymbol{K}^{\ }_{-}$) may be induced by Coulomb or electron-phonon
interactions.  Many theoretical proposals to understand these new
plateaus and predict new ones have been made
(for a review, see Ref.~\onlinecite{graphene-review}).

Here, we shall focus on the particularly interesting $n=0$ Landau
level.  The dependence of the longitudinal resistance $R^{\ }_{xx}$ at
this level (fixing the chemical potential $\mu=0$) for graphene
deposited on a Si-SiO$^{\ }_2$ substrate under intense uniform
magnetic fields $B$ of up to 32 [T] has been measured in
Refs.~\onlinecite{Checkelsky08} and \onlinecite{Checkelsky09}.  At
fixed low temperatures (0.3--5[K]), $R^{\ }_{xx}$ grows exponentially
with magnetic fields $B>17$ [T] upon approaching a sample-dependent
critical field strength $B^{\ }_{c}$.  The critical field $B^{\ }_{c}$
is larger for samples with lower zero-field mobility $\mu^{\
}_{e}$. There also appears to be a scaling regime with the scaling
function $R(b)=c^{\ }_{1}\exp[2\times c^{\
}_{2}(1-b)^{-\varrho}]\;\Omega/\square$ with $c^{\ }_{1}=440$, $c^{\
}_{2}=1.54$, $b=B/B^{\ }_{c}(\mu^{\ }_{e})<1$, and $\varrho=1/2$
fitting well the magnetic field dependence of $R^{\ }_{xx}$ for
samples characterized by the mobility-dependent $B^{\ }_{c}$.  This
scaling is reminiscent of the Kosterlitz-Thouless (KT) divergence of
the correlation length in the 2D \textit{classical} $XY$
model.~\cite{Kosterlitz73} Finally, the temperature dependence of
$R^{\ }_{xx}$ for a given sample with a fixed $B$ very close to $B^{\
}_{c}$ approaches the thermally activated form 
$\exp(2\Delta^{\ }_{\mathrm{a}}/T)$ 
on the interval 2~K$<T<$16~K.  ~\cite{footnote: exp Delta}

Our proposal to explain this KT scaling is closely related to the
scenario from Ref.~\onlinecite{Nomura09}. However, the disorder
was treated at the level of the self-consistent Born approximation in
Ref.~\onlinecite{Nomura09}, while the role of the Coulomb interaction,
within the Hartree-Fock approximation, was emphasized.  
Here, we show that the electron-phonon interaction is critical
to the selection of  the Kekul\'e instability among all 15
instabilities that can open a gap at the Dirac point
while preserving the electron charge as a good quantum number.%
~\cite{Ryu09} Moreover, we argue that a non-perturbative
treatment of the disorder is also \textit{essential} to explain the
striking fact that $B^{\ }_{c}(\mu^{\ }_{e})$ is independent of $T$ at
low temperatures: it is the 2D classical random phase $XY$ model that
describes the phase transition observed in
Ref.~\onlinecite{Checkelsky09}. Within this classical random phase
$XY$ model, vortices unbind when the magnetic field is below 
$B^{\ }_{c}$, 
and thus deconfined fractionally charged quasiparticles can be
sustained in graphene.

The line of arguments in this paper can be summarized as follows.
For the relevant range of magnetic fields, the dominant energy scale
originates from the kinetic energy modified by the orbital coupling
to the magnetic field. The Coulomb interaction on the length scale
of the magnetic length -- the long-range Coulomb interaction -- 
is the leading subdominant energy scale.
To these leading and subleading orders, 
the dynamics of  interacting electrons in graphene preserves 
a combined U(4) symmetry arising from the conservation of
the electron charge, axial charge,~\cite{Jackiw07,Ryu09} and two independent
spin-1/2 and valley-1/2 SU(2) symmetries.
This continuous symmetry can only be broken spontaneously
at zero temperature according to the Mermin-Wagner theorem.
Any non-vanishing ordering temperature thus requires
the explicit breaking of this U(4) symmetry
down to products of the U(1) subgroup enforcing the conservation
of the electron quantum number with either finite subgroups
or U(1) subgroups of SU(4). 
The bare Zeeman, electron-phonon, and short-range Coulomb interactions, 
although close to two orders of magnitude smaller than
the long-range Coulomb interaction, provide these anisotropies.
In particular, the electron-phonon interaction favors
the Kekul\'e instability over other instabilities, for example magnetic ones,
so that, even if one could turn off electron-electron interactions, 
the Kekul\'e instability would still compete with the Zeeman interaction. 
Hence, we first start by considering the effects of lattice
distortions in the case of noninteracting electrons
and without the Zeeman interaction in 
Sec.~\ref{sec: The Kekule instability}.
In this approximation, we show that 
the Kekul\'e instability triggered by the electron-phonon coupling
can explain the experiments of Ref.~\onlinecite{Checkelsky09}.  
In Sec.~\ref{sec: Competing Zeeman energy}, we show that as
long as the Kekul\'e energy gap in the absence of the Zeeman coupling
is greater than twice the Zeeman splitting (once that is turned on), 
the results of Sec.~\ref{sec: The Kekule instability} remain unchanged. 
However, for graphene, the value we obtain theoretically for the Kekul\'e gap 
using the electron-phonon coupling alone is too close to this threshold
of twice 
the Zeeman splitting (as reported in Ref.~\onlinecite{Jiang07}).
Therefore, a proper treatment of the electron-electron interactions 
is needed for selecting the correct low-temperature phase
in Ref.~\onlinecite{Checkelsky09}.
We restore the full Coulomb electron-electron interactions in 
Sec.~\ref{sec: Electron-electron interactions}. 
We first argue that
the lattice nearest-neighbor Coulomb interaction boosts the
electron-phonon coupling and stabilizes the Kekul\'e order in spite
of the competing Zeeman splitting.
We then argue that the KekulŽ instability and the Zeeman splitting provide the needed anisotropies to reduce the symmetry and evade the Mermin-Wagner theorem that precludes ordering at a finite temperature. The finite ordering temperature is shown to depend on both the SU(4) symmetric electron-electron interactions and on the explicit anisotropies, including those from the electron-phonon interactions.
We conclude that the results obtained within the approximations of
Sec.~\ref{sec: The Kekule instability} remain valid, 
although the temperature scale for the transition is increased.

\section{
The Kekul\'e instability
        }
\label{sec: The Kekule instability}

In this section we are going to ignore the Zeeman 
and electron-electron interactions altogether.
We will revisit this approximation 
in Secs.~\ref{sec: Competing Zeeman energy}
and~\ref{sec: Electron-electron interactions}.
We shall first derive the mean-field Kekul\'e instability
induced by phonons in the presence of a magnetic field. We shall then prove the existence of zero energy states when the Kekul\'e order supports a defect in the form of a vortex in the presence of a uniform magnetic field. 

\subsection{
The mean-field Kekul\'e order.
           }
           
In the continuum limit,
the orbital contribution from a magnetic field
$\boldsymbol{B}=\boldsymbol{\nabla}\wedge\boldsymbol{A}$
to pristine graphene reads 
\begin{subequations}
\label{eq: def rep continuum limit}
\begin{equation}
\mathcal{H}^{\ }_{0}=
\hbar
v^{\ }_{\mathrm{F}}
\left(\boldsymbol{k}-\frac{e}{\hbar c}\boldsymbol{A}\right)
\cdot
\boldsymbol{\alpha}
\label{eq:def-mathcal{H}0}
\end{equation}
where the 2D wave number $\boldsymbol{k}$ has the 
components $k^{\ }_{1}$ and $k^{\ }_{2}$, 
and the Dirac matrices are
\begin{equation}
\alpha^{\ }_{1}=
\sigma^{\ }_{3}\otimes 
\tau^{\ }_{1}\otimes
s^{\ }_{0},
\qquad
\alpha^{\ }_{2}=
\sigma^{\ }_{3}\otimes 
\tau^{\ }_{2}\otimes
s^{\ }_{0}.
\end{equation}
\end{subequations}
The unit $2\times 2$ matrices 
$\sigma^{\ }_{0}$, 
$\tau^{\ }_{0}$,
and 
$s^{\ }_{0}$
together with the Pauli matrices 
$\boldsymbol{\sigma}$,
$\boldsymbol{\tau}$,  
and 
$\boldsymbol{s}$
act on the valley-1/2 
($\boldsymbol{K}^{\ }_{+}$ and $\boldsymbol{K}^{\ }_{-}$),
sublattice-1/2  
(A and B),
and
spin-1/2 ($\uparrow$ and $\downarrow$)
2-dimensional subspaces of graphene, respectively.

The spectrum of $\mathcal{H}^{\ }_{0}$ is 4-fold degenerate,
for $\mathcal{H}^{\ }_{0}$ commutes with the 16 Hermitean generators 
\begin{subequations}
\label{eq: generators U(4) sym H0}
\begin{eqnarray}
X^{\ }_{00\kappa}:=
\sigma^{\ }_{0}
\otimes
\tau^{\ }_{0}
\otimes
s^{\ }_{\kappa},
&&
X^{\ }_{13\kappa}:=
\sigma^{\ }_{1}
\otimes
\tau^{\ }_{3}
\otimes
s^{\ }_{\kappa},
\label{eq: generators U(4) sym H0-1}
\\
X^{\ }_{23\kappa}:=
\sigma^{\ }_{2}
\otimes
\tau^{\ }_{3}
\otimes
s^{\ }_{\kappa},
&&
X^{\ }_{30\kappa}:=
\sigma^{\ }_{3}
\otimes
\tau^{\ }_{0}
\otimes
s^{\ }_{\kappa},
\label{eq: generators U(4) sym H0-2}
\end{eqnarray}
(here $\kappa=0,1,2,3$) of the Lie group
\begin{equation}
\mathrm{U(4)}=
\mathrm{U(1)}
\times
\mathrm{SU(4)}
\simeq
\mathrm{SO(6)}.
\label{eq: Lie group symmetry H0}
\end{equation} 
\end{subequations}
Notice the electron charge U(1) subgroup generated by
\begin{subequations}
\begin{equation}
X^{\ }_{000},
\label{eq: U(1) charge}
\end{equation}
the spin-1/2 SU(2) subgroup generated by
\begin{equation}
X^{\ }_{001},\qquad
X^{\ }_{002},\qquad
X^{\ }_{003},
\label{eq: U(2) charge+spin symmetry}
\end{equation}
and the valley-1/2 SU(2) subgroup generated by
\begin{equation}
X^{\ }_{130},\qquad
X^{\ }_{230},\qquad
X^{\ }_{300}\equiv\gamma^{\ }_{5}.
\label{eq: U(2) charge+valley symmetry}
\end{equation}
\end{subequations}

A Kekul\'e instability is 
a periodic modulation of the nearest-neighbor hopping
amplitude in graphene with the wave vector 
$\boldsymbol{K}^{\ }_{+}-\boldsymbol{K}^{\ }_{-}$.~\cite{Hou07}
In the continuum approximation~(\ref{eq: def rep continuum limit}), 
it is represented by the Hermitean mass matrix
\begin{equation}
\mathcal{M}^{\ }_{\mathrm{K}}:=
\boldsymbol{\Delta}
\cdot
\boldsymbol{M}
\equiv
\Delta^{\ }_{1}
M^{\ }_{1}
+
\Delta^{\ }_{2}
M^{\ }_{2}
\end{equation}
that is parametrized by the real-valued numbers
$\Delta^{\ }_{1}$
and
$\Delta^{\ }_{2}$.~\cite{Hou07}
The Kekul\'e mass matrices are
\begin{equation}
M^{\ }_{1}:= 
\sigma^{\ }_{1}\otimes 
\tau^{\ }_{0} \otimes
s^{\ }_{0},
\qquad
M^{\ }_{2}:=
-\sigma^{\ }_{2}\otimes 
\tau^{\ }_{0} \otimes
s^{\ }_{0},
\end{equation}
where a mass matrix is any Hermitean matrix
\begin{equation}
X^{\ }_{\kappa^{\ }_{1}\kappa^{\ }_{2}\kappa^{\ }_{3}}:= 
\sigma^{\ }_{\kappa^{\ }_{1}}
\otimes
\tau^{\ }_{\kappa^{\ }_{2}}
\otimes
s^{\ }_{\kappa^{\ }_{3}},
\qquad
\kappa^{\ }_{1},\kappa^{\ }_{2},\kappa^{\ }_{3}=0,1,2,3,
\end{equation}
that anticommutes with $\mathcal{H}^{\ }_{0}$.
[There are 16 mass matrices as shown in Ref.~\onlinecite{Ryu09}.
They generate a group U(4) distinct from the group U(4) whose
generators are given by Eq.~(\ref{eq: generators U(4) sym H0}).]
In the continuum approximation, 
only the 8 generators 
\begin{equation}
X^{\ }_{00\kappa},
\qquad
\Delta^{\ }_{1} X^{\ }_{13\kappa}
-
\Delta^{\ }_{2} X^{\ }_{23\kappa},
\label{eq: generators that commute with Kekule}
\end{equation}
(where $\kappa=0,1,2,3$) 
of the Lie group~(\ref{eq: generators U(4) sym H0})
commute with
\begin{equation}
\mathcal{H}:=
\mathcal{H}^{\ }_{0}
+
\mathcal{M}^{\ }_{\mathrm{K}}.
\label{eq:def-mathcal{H}}
\end{equation}

In the presence of the uniform magnetic field 
$B=\partial^{\ }_{1}A^{\ }_{2}- \partial^{\ }_{2}A^{\ }_{1}$
and the uniform  Kekul\'e order parameter
$\Delta= \Delta^{\ }_{1} + {i} \Delta^{\ }_{2}= \Delta^{\ }_{0}
e^{{i}\theta^{\ }_{0}}$, for 
$0\leq\Delta^{\ }_{0}$ and $0\leq\theta^{\ }_{0}<2\pi$
the single-particle spectrum is the shifted Landau spectrum
\begin{equation}
\varepsilon^{\ }_{N,\pm}=
\pm
\sqrt{
\left(\hbar\omega^{\ }_{c}\right)^{2}N+\Delta^{2}_{0}
     }
\label{eq:Landau-spectrum}
\end{equation}
where $N=0,1,2,\cdots$. 

The mean-field value of the Kekul\'e gap $2 \Delta^{\ }_{0}$
induced by a magnetic field
and by an electron-phonon coupling in the single-particle
approximation (see also Ref.~\onlinecite{Ajiki95}) is obtained by
balancing the gains in the electronic energy against 
the losses in the elastic energy of the lattice:
\begin{subequations}
\label{eq: energetic balance}
\begin{equation}
\delta E\equiv
\delta E^{\ }_{\mathrm{elec}}
+
\delta E^{\ }_{\mathrm{pho}}
\end{equation}
where
\begin{equation}
\begin{split}
\delta E^{\ }_{\mathrm{elec}}=&\,
2\times
\frac{\mathcal{A}}{2\pi\ell^{2}_{B}}
\sum_{N=0}^{\infty}
\left(
2
-
\delta^{\ }_{0,N}
\right)
\left(
\varepsilon^{\ }_{N,-}
-
\varepsilon^{\ }_{-N}
\right)
\\
=&\,
-
2\times
\frac{\mathcal{A}}{2\pi\ell^{2}_{B}}
\Delta^{\ }_{0}
+
\mathcal{O}
\left(
\frac{
\Delta^{2}_{0}
     }
     {
\hbar\omega^{\ }_{\mathrm{c}}
     }
\right)
\end{split}
\label{eq: electronic gain}
\end{equation}
is the electronic gain in energy, while 
\begin{equation}
\delta E^{\ }_{\mathrm{pho}}=
\frac{\mathcal{N}}{2}
K^{\ }_{\Delta}
u^{2}_{0}
=
\frac{\mathcal{N}}{2}
\frac{K^{\ }_{\Delta}}{f^{2}_{\Delta}}
\Delta^{2}_{0}
\label{eq: elastic loss}
\end{equation}
\end{subequations}
is the cost in elastic energy. Here, $K^{\ }_{\Delta}$ is the bond
elastic constant and $f^{\ }_{\Delta}$ 
ties the gap $ 2 \Delta^{\ }_{0}$ to the
atomic displacement $u^{\ }_{0}$ for the Kekul\'e order, which results
because of the change in hopping matrix elements from the orbital
overlaps. (The contribution to the electronic gain from the levels
$N=1,2,\cdots$ can be absorbed into a downward renormalization of the
elastic rigidity $K^{\ }_{\Delta}/f^{2}_{\Delta}$.~\cite{Fuchs07}) We
have assumed that the cyclotron energy is the largest energy scale of
the problem. We have also introduced the area $\mathcal{A}$ and the
number $\mathcal{N}=4\mathcal{A}/(3\sqrt{3}\,\mathfrak{a}^{2})$ of
sites of graphene ($\mathfrak{a}\approx1.42$[\r A] is the lattice
spacing). The absolute minimum of Eq.~(\ref{eq: energetic balance})
is obtained for the mean-field value
\begin{equation}
\Delta^{\ }_{0}=
\frac{f^{2}_{\Delta}}{\mathcal{N}K^{\ }_{\Delta}}
\frac{\mathcal{A}}{\pi\ell^{2}_{B}}=
\frac{3\sqrt{3}\,\mathfrak{a}^{2}}{4\pi\ell^{2}_{B}}
\frac{f^{2}_{\Delta}}{K^{\ }_{\Delta}}
\label{eq: mean-field sol}
\end{equation}
of the Kekul\'e single-particle mass.
If, following Ref.~\onlinecite{Ajiki95}, we make the estimate
$f^{2}_{\Delta}/K^{\ }_{\Delta}\approx 6.366$ [eV], 
then the mean-field Kekul\'e gap is estimated to be
\begin{equation}
2\times\Delta^{\ }_{0}\approx
1.86 \times B\hbox{[K]},
\label{eq:estimate-Kekule}
\end{equation}
in units of Kelvin
(with $B$ measured in units of the Tesla).
The single-particle electronic
gap~(\ref{eq:estimate-Kekule})
overestimates by a factor of $\approx 2$ the measured 
activation gaps.%
~\cite{Checkelsky09,footnote: exp Delta}

We now discuss the all important phase fluctuations
of the Kekul\'e order parameter and the effective action that governs them. 
The complex-valued  
Kekul\'e order parameter $\Delta=\Delta_0\,e^{i\theta}$ has a
phase $\theta(\boldsymbol{r},t)$ that can fluctuate in space 
and in time. The dependence of the effective action 
on this phase stems from the dynamics of the phonons 
on the one-hand and from integrating out the electronic 
degrees of freedom on the other hand.

In addition, we shall also account for the fluctuations
in the hopping matrix elements due to disorder or to local
curvature effects (tiny ripples) on the surface of graphene,
\begin{equation}
\delta\mathcal{H}=
-\hbar
v^{\ }_{\mathrm{F}}\,
\boldsymbol{A}^{\ }_{5}\,
\cdot
\gamma^{\ }_{5}\, 
\boldsymbol{\alpha}.
\label{eq:delta-H5}
\end{equation}
The axial vector potential $\boldsymbol{A}^{\ }_{5}$ 
encodes these disorder effects in the hopping matrix
elements. Notice the 
\begin{equation}
\gamma^{\ }_{5}=
\sigma^{\ }_{3}\otimes
\tau^{\ }_{0}\otimes 
s^{\ }_{0}
\end{equation} 
matrix, so that $\boldsymbol{A}^{\ }_{5}$ 
couples with opposite signs to the two flavors at
$\boldsymbol{K}^{\ }_{+}$ and $\boldsymbol{K}^{\ }_{-}$, 
in contrast to the electromagnetic vector potential 
$\boldsymbol{A}$.  

As shown by Jackiw and Pi,%
~\cite{Jackiw07} 
this axial vector potential is intimately
related to the phase of the Kekul\'e order parameter.
The axial gauge transformation
$
\Psi\to 
e^{i\gamma^{\ }_{5}\,\theta/2}
\;
\Psi
$ 
on the single-particle spinor $\Psi$ for electrons, 
removes the phase of the Kekul\'e order parameter,
$\Delta\to \Delta\,e^{-i\theta}=\Delta^{\ }_{0}$, 
by shifting
${A^{\ }_{5}}_\mu\to{A^{\ }_{5}}_\mu-\partial^{\ }_{\mu}\theta/2$ 
(where $\mu=0,1,2$). The
effective action that follows
upon integrating the electrons can be inferred from symmetry.%
~\cite{Ryu09}
The leading term is, at $T=0$,
[$\partial^{\ }_{\mu}\equiv
(v^{-1}_{\mathrm{F}}\partial^{\ }_{t},
 \partial^{\ }_{\boldsymbol{r}})$]
\begin{equation}
S=
\frac{J}{2\hbar}\int dt\;d^2r 
\left(\partial^{\ }_{\mu}\theta-2A^{\ }_{5\mu}\right)^2.
\label{eq:random-XY-action}
\end{equation}
The stiffness $J$ was computed at zero magnetic field in
Ref.~\onlinecite{Ryu09}. Now, we compute $J$ in the presence of
$B$. To this end, we consider the time independent Kekul\'e texture 
with $\theta=\boldsymbol{q}\cdot\boldsymbol{r}$ 
and $\boldsymbol{A}^{\ }_{5}=0$, 
or the gauge equivalent 
$\theta=0$ and $\boldsymbol{A}^{\ }_{5}=-\boldsymbol{q}/2$. 
We solve for the eigenvalues of 
$\mathcal{H}+\delta\mathcal{H}$
treating $\boldsymbol{A}^{\ }_{5}=-\boldsymbol{q}/2$
to second order in perturbation theory
(see details in Appendix \ref{sec:Stiffness}).
The electronic energy cost of 
$\boldsymbol{A}^{\ }_{5}=-\boldsymbol{q}/2$
to order $|\boldsymbol{q}|^2$
determines the stiffness
\begin{equation}
J=
\frac{\Delta^{\ }_{0}}{2 \pi}.
\label{eq:estimate-for-J} 
\end{equation}
(This result coincides with the stiffness computed at zero field in
Ref.~\onlinecite{Ryu09}, once spin degeneracy is accounted there).

We shall treat the $\boldsymbol{A}^{\ }_{5}$ in
Eq.~(\ref{eq:random-XY-action}) as a static random phase on
phenomenological grounds dictated by symmetry. A microscopic
justification for this step is the following. In a realistic sample,
the matrix hopping elements between nearest-neighbor atoms will not be
all the same, but there will be fluctuations caused by several factors,
strain when in contact with the substrate being one factor. The random
hopping amplitudes not only cause the phase in
Eq.~(\ref{eq:random-XY-action}) to be random, but also the coupling
$J$ to be spatially varying. However, weak random exchange is
irrelevant, and we thus simply focus on the uniform $J$ case.

The effective action on length scales larger than $\ell^{\ }_{B}$ is
thus that of the quantum random phase $XY$ model in 2D whereby we
assume the white-noise distribution of vanishing mean (overline
denotes disorder averaging)
\begin{equation}
\overline{
A^{\ }_{5i}(\boldsymbol{r})
A^{\ }_{5j}(\boldsymbol{r}')
         }= 
g\,
\delta^{\ }_{ij}\,
\delta^{(2)}
\left(\boldsymbol{r}-\boldsymbol{r}'\right),
\quad i,j=1,2.
\label{eq: disorder correlation}
\end{equation}
It is crucial to appreciate that
$g$ is a dimensionless coupling constant. 
It thus depends on the magnetic field through a function 
$g(\ell/\mathfrak{a},\ell^{\ }_{B}/\ell)$ of two arguments. The argument
$\ell/\mathfrak{a}$ is the mean free path 
associated with the disorder for the system at $B=0$
in units of the lattice spacing $\mathfrak{a}$. 
The argument $\ell^{\ }_{B}/\ell$ must also be present
because as $\ell^{\ }_{B}$ decreases with increasing $B$, 
the single-particle overlaps due to disorder of (otherwise orthogonal) 
orbitals decrease.

Next, we are going to argue that 
the 2D random phase $XY$ model model defined by Eqs.%
~(\ref{eq:random-XY-action})-(\ref{eq: disorder correlation})
solves two experimental puzzles 
observed in Ref.~\onlinecite{Checkelsky09}. 
Experimentally, the resistivity of graphene is fit
to a form consistent with the scaling of the correlation length 
$\xi$ of the 2D \textit{classical} $XY$ model, where the distance to the
critical point is measured in terms of the reduced field
$b=B/B^{\ }_{c}$. Why is it that the data fits a transition
as a function of $b$
in the 2D universality class of the classical as opposed to
the quantum $XY$ model? Why is it that the
resistivities are rather temperature independent for temperatures
differing by a factor of 5, from 0.3 K to 1.5 K? 

First, the measured KT scaling could be explained if the
fluctuations driving the transition were 1D but of quantum origin. 
For instance, if the transition had to do with
edge physics. This idea has been proposed in
Ref.~\onlinecite{Shimshoni09}. 
The problem with this scenario is that it requires 
1) a large Zeeman coupling, 
2) a critical density of magnetic impurities,
and 3) the ordered phase in a 1D quantum system does not survive
finite temperatures so that the size of the region in which the
correlation length scales in a KT fashion near the quantum
critical point is bounded from above by an exponentially small
crossover temperature.

Second, a quantum critical point of the 2D quantum $XY$
model~(\ref{eq:random-XY-action}) \textit{without static random
phases} does not obey the KT scaling that fits the data in
Ref.~\onlinecite{Checkelsky09} over 3 decades of values of
longitudinal resistances $R(b)$.  On the other hand, a KT scaling is
compatible with the clean limit of model~(\ref{eq:random-XY-action})
if the relevant range of temperatures for which quasi-long-range order
holds is large enough so that quantum critical fluctuations can be
neglected (see Ref.~\onlinecite{Sachdev99}). However, even if the
quantum critical fluctuations 
can be neglected when the temperature changes
from 0.3 K to 1.5 K, 
the position of the critical field $B^{\ }_{c}$
should move with temperature, for the location of the clean KT
transition is where $J(B)=2T/\pi$.

\begin{figure}
\includegraphics[angle=0,scale=0.75]{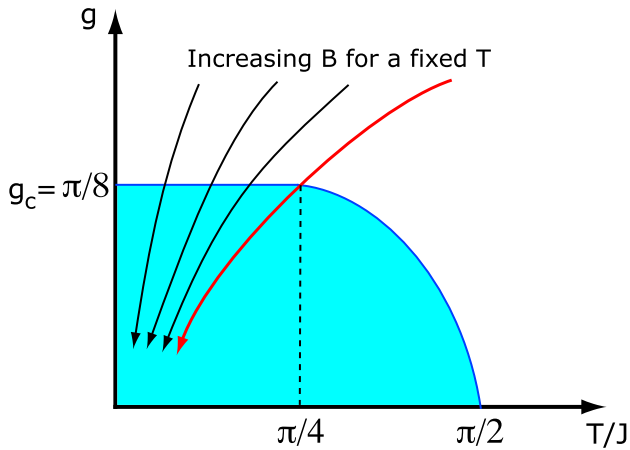}
\caption{
(Color online)
Phase diagram of the 2D classical random phase $XY$ model
after Ref.~\onlinecite{Korshunov06}.
The horizontal axis represents the temperature $T$ in units
of the stiffness $J$.
The vertical axis represents the dimensionless disorder
strength $g$ 
(the variance for white-noise and Gaussian correlated random phases).
The shaded region depicts the quasi-long-range ordered phase
for which the interaction between vortices is logarithmic.
The complementary white region depicts the paramagnetic phase for which
the interaction between vortices is screened beyond the screening
length $\xi$.
At the transition line that separates the two phases, 
vortices undergo a confining-deconfining transition. 
We interpret the divergence of the resistance of graphene
subjected to a strong increasing magnetic field $B$ measured
in Ref.~\onlinecite{Checkelsky09}
as being governed by the increase of the screening length $\xi(B)$
in the paramagnetic region along a RG trajectory parametrized by $B$.
Here, $T/J$ and $g$ are both decreasing functions of $B$.
        }
\label{fig:2Dphase-deiagram}
\end{figure}

Third, the disorder in Eq.~(\ref{eq:random-XY-action}) 
can explain the observed KT scaling.
Indeed, the 2D classical random phase $XY$ model has the
remarkable feature that its boundary separating the
quasi-long-range ordered phase from the paramagnetic phase runs
(nearly) parallel to the temperature axis when 
$T<T^{\ }_{\mathrm{KT}}/2$, where 
$T^{\ }_{\mathrm{KT}}=\pi/2\;J$.~\cite{Korshunov06} (This feature is
non-perturbative in disorder.~\cite{Mudry99,Carpentier00}) However, no
consensus has emerged on the exponent $\varrho$ that characterizes the
diverging correlation length $\exp(b/x^{\varrho})$ upon approaching the
boundary (it is debated if it is $\varrho=1/2$ or $1$). The phase
diagram is shown in Fig.~\ref{fig:2Dphase-deiagram}. Also shown 
are the trajectories of coordinates
$\left(\;T/J(B),g(\ell/\mathfrak{a},\ell^{\ }_{B}/\ell)\;\right)$ as
the magnetic field $B$ is changed for fixed $\ell$ and
$T$. When these trajectories enter the quasi-long-ordered
phase through the disorder-controlled boundary parallel to the $T/J$ axis, one has the condition
\begin{subequations}
\begin{equation}
g(\ell/\mathfrak{a},\ell^{\ }_{B}/\ell)=\pi/8, 
\hbox{ i.e., }
B=B^{\ }_{c}(\mathfrak{a},\ell/\mathfrak{a})
\end{equation}
which is temperature independent as long as
\begin{equation}
T<T^{\ }_{\mathrm{KT}}/2,
\hbox{ i.e., }
T<\Delta^{\ }_{0}(B^{\ }_{c})/8.
\label{eq:Tmax}
\end{equation}
\end{subequations}
One can estimate the upper range of temperatures 
for which the transition is KT-like as a function of the scaling
parameter $b=B/B^{\ }_{c}$ used in Ref.~\onlinecite{Checkelsky09}.
For $B^{\ }_{c}\approx 29.1$ [T], 
one finds using Eqs.~(\ref{eq:estimate-Kekule})
and~(\ref{eq:Tmax}) that the 2D classical scaling as a
function of $b$ holds as long as $T<3.5$ K.

These results thus explain rather well the features observed in
Ref.~\onlinecite{Checkelsky09}. Accordingly, the Kekul\'e
distortion, which does not appear to occur in graphene at zero
magnetic field, is induced by the presence of the magnetic field above a
critical field $B^{\ }_{c}$ that depends on the
disorder strength and is thus sample dependent. 

\subsection{
Zero modes in a uniform magnetic field
           }

One of the surprising features in graphene is that
vortices in the Kekul\'e dimerization pattern bind fractionalized
excitations when $B=0$.~\cite{Hou07} 
We now show analytically that fractionalization is robust
to the presence of $B$, i.e., that there is a zero mode due to vortices
that binds a fraction of the electron for any value of $B$.

\begin{figure}
\begin{center}
\includegraphics[angle=0,scale=0.95]{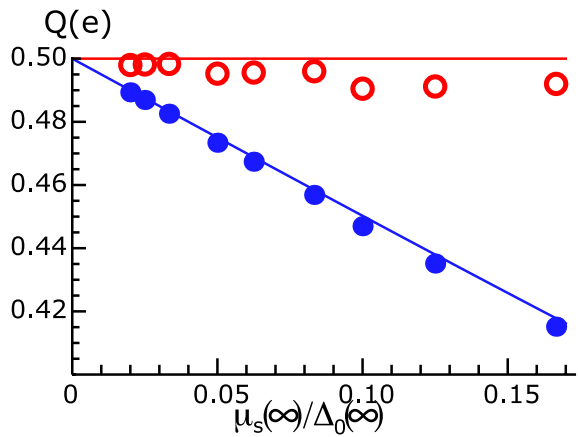}
\end{center}
\caption{
Fractionalized charge $Q$ in units of the electric charge $e$ 
of electrons in the presence of a magnetic field $B=20$ (T)
as a function of the ratio between the staggered chemical
potential $\mu^{\ }_{\mathrm{s}}(\infty)$
and the Kekul\'e gap 
$\Delta^{\ }_{0}(\infty)$
very far away from a single vortex (filled circles)
or very far away from a pair of topological defects made of
a vortex and a half axial flux (open circles).
The lattice model consists of $113\times 80$ sites. 
The solid lines are the analytical results
presented in Ref.~\onlinecite{Ryu09} when $B=0$.
        }
\label{fig:charge}
\end{figure}

We thus seek $\varepsilon=0$ solutions of the Dirac operator%
~(\ref{eq:def-mathcal{H}}) when the Kekul\'e order parameter
supports the vortex
\begin{equation}
\Delta(\boldsymbol{r})\equiv
\Delta(\rho,\theta)=
\Delta^{\ }_{0}
e^{{i}\theta^{\ }_{0}}
e^{{i}n_{\rm v}\theta}.
\end{equation}
To this end, we use polar coordinates and the symmetric
gauge $\boldsymbol{A}=B(-y,x)/2$.  Without loss of generality, we
restrict the vorticities to $n_{\rm v}=\pm1$. It follows from the sublattice
symmetry in the problem that, as in the case when $B=0$, the zero-mode
solutions have support on either the A (for $n_{\rm v}=-1$) or B sublattice
(for $n_{\rm v}=+1$). The solutions in the presence of the magnetic field 
(see details in Appendix \ref{sec:zero-mode})
are
\begin{equation}
\label{eq: midgap states}
\Psi^{\ }_{\mathrm{A}}(\boldsymbol{r})=
\begin{pmatrix}
0
\\
u(\rho)
\\
v(\rho)
\\
0
\end{pmatrix},
\qquad
\Psi^{\ }_{\mathrm{B}}(\boldsymbol{r})=
\begin{pmatrix}
v(\rho)
\\
0
\\
0
\\
-u(\rho)
\end{pmatrix},
\end{equation}
where
\begin{subequations}
\begin{eqnarray}
u(\rho)&=&{i}\,e^{-{i}\theta^{\ }_{0}/2}
\sqrt{p}\;
D^{\ }_{
-
p
-
1 
       }(\rho/\ell^{\ }_{B}),
\label{eq:u-and-v1}
\\
v(\rho)&=&e^{+{i}\theta^{\ }_{0}/2}
D^{\ }_{
-
p
       }(\rho/\ell^{\ }_{B})
\;,
\label{eq:u-and-v2}
\end{eqnarray}
$D^{\ }_{p}(z)$ is the parabolic cylinder
function,~\cite{Gradshteyn00} and the dimensionless ratio
\begin{equation}
p=
\frac{
2\Delta^{2}_{0}
     }
     {
(\hbar\omega^{\ }_{\mathrm{c}})^{2}
     }.
\end{equation}
\end{subequations}
The existence of a single midgap state for a vortex
($n_{\rm v}=+1$) or anti-vortex ($n_{\rm v}=-1$) implies that these topological
defects bind half of the electron charge $\pm1/2$.~\cite{Hou07}
We have verified numerically 
(see Fig.~\ref{fig:charge})
that the effect of
breaking the sublattice symmetry with a staggered chemical potential
$\mu^{\ }_{\mathrm{s}}$ (a coexisting charge density wave) is to move
the midgap states away from $\mu=0$.
According to Fig.~\ref{fig:charge},
the fractional charge bound to a vortex 
acquires a dependence on the value of 
$\mu^{\ }_{\mathrm{s}}$. Remarkably,
we can fit this dependence in the presence of the strong magnetic field
$B$ with the results obtained in Refs.%
~\onlinecite{Hou07},
\onlinecite{Jackiw07},
\onlinecite{Chamon08a},
\onlinecite{Chamon08b},
and
\onlinecite{Ryu09}
in the absence of the magnetic field!
Altogether, our analytical and numerical results show that the
fractionalization of the electric charge of electrons is robust 
to any magnetic field.

\section{
Competing Zeeman energy
        }
\label{sec: Competing Zeeman energy}

So far, we have ignored the Zeeman energy.
The reason to do so is two-fold.
First, the Kekul\'e order parameter is
a spin singlet and thus all preceding considerations
apply in the presence of a Zeeman interaction,
if it can be established that the Kekul\'e instability
survives the presence of the Zeeman interaction.
Second, as we show below, the mean-field Kekul\'e
instability~(\ref{eq: mean-field sol}) 
is unchanged by the addition to $\mathcal{H}$
defined in Eq.~(\ref{eq:def-mathcal{H}})
of the Zeeman interaction
\begin{equation}
\mathcal{H}^{\ }_{\mathrm{Z}}:=
\Delta^{\ }_{\mathrm{Z}}\,
\sigma^{\ }_{0}\otimes
\tau^{\ }_{0}\otimes
s^{\ }_{3}
\label{eq: def Zeeman interaction}
\end{equation}
if 
\begin{equation}
0\leq \Delta^{\ }_{\mathrm{Z}}<\frac{\Delta^{\ }_{0}}{2},
\label{eq: condition to drop Zeeman}
\end{equation}
while there is no Kekul\'e instability if
\begin{equation}
0\leq \frac{\Delta^{\ }_{0}}{2}<\Delta^{\ }_{\mathrm{Z}};
\label{eq: condition to not drop Zeeman}
\end{equation}
the case of 
\begin{equation}
\frac{\Delta^{\ }_{0}}{2}=
\Delta^{\ }_{\mathrm{Z}}
\label{eq: 1st order phase transition}
\end{equation}
being a mean-field first-order transition.
Hence, if condition~(\ref{eq: condition to drop Zeeman}) holds, 
it is legitimate to ignore the Zeeman energy altogether
when computing the magnitude of the Kekul\'e gap.

This is not to say that the Zeeman interaction
does not play an essential role,
for it is it together with the Kekul\'e order parameter
that establishes the desired pattern of explicit symmetry breaking
\begin{subequations}
\label{eq: desired pattern symmetry breaking}
\begin{eqnarray}
&&
\mathcal{H}^{\ }_{0}\to
\mathcal{H}^{\ }_{0}
+
\mathcal{H}^{\ }_{\mathrm{K}}
+
\mathcal{H}^{\ }_{\mathrm{Z}},
\\
&&
\mathrm{U}(4)\to
\mathrm{U}(1)
\times
\mathrm{U}(1)
\times
\mathrm{U}(1)
\times
\mathrm{U}(1)
\end{eqnarray}
with the residual unbroken U(1) symmetry generators
\begin{equation}
X^{\ }_{000},
\
X^{\ }_{003},
\
\Delta^{\ }_{1} X^{\ }_{130}
-
\Delta^{\ }_{2} X^{\ }_{230},
\
\Delta^{\ }_{1} X^{\ }_{133}
-
\Delta^{\ }_{2} X^{\ }_{233},
\label{eq: generators that commute with Kekule+Zeeman}
\end{equation}
\end{subequations}
respectively.

To understand results
(\ref{eq: condition to drop Zeeman})-(\ref{eq: 1st order phase transition}), 
it suffices to observe that
the presence of the Zeeman interaction%
~(\ref{eq: def Zeeman interaction})
changes the single-particle spectrum~(\ref{eq:Landau-spectrum})
to
\begin{equation}
\varepsilon^{\ }_{N,\pm}\to
\left\{
\begin{array}{l}
+
\sqrt{
\left(\hbar\omega^{\ }_{c}\right)^{2}N+\Delta^{2}_{0}
     }
+
\Delta^{\ }_{\mathrm{Z}},
\\
+
\sqrt{
\left(\hbar\omega^{\ }_{c}\right)^{2}N+\Delta^{2}_{0}
     }
-
\Delta^{\ }_{\mathrm{Z}},
\\
-
\sqrt{
\left(\hbar\omega^{\ }_{c}\right)^{2}N+\Delta^{2}_{0}
     }
+
\Delta^{\ }_{\mathrm{Z}},
\\
-
\sqrt{
\left(\hbar\omega^{\ }_{c}\right)^{2}N+\Delta^{2}_{0}
     }
-
\Delta^{\ }_{\mathrm{Z}},
\end{array}
\right.
\label{eq:Landau-spectrum with Zeeman}
\end{equation}
where $N=0,1,2,\cdots$.
It then follows that the gain in the kinetic energy%
~(\ref{eq: electronic gain})
is changed to
\begin{equation}
\begin{split}
\delta E^{\ }_{\mathrm{elec}}\to&\,
-
2\times
\frac{\mathcal{A}}{2\pi\ell^{2}_{B}}
\Theta\left(\Delta^{\ }_{0}-\Delta^{\ }_{\mathrm{Z}}\right)
\left(
\Delta^{\ }_{0}-\Delta^{\ }_{\mathrm{Z}}
\right)
\\
&\,+
4\times
\frac{\mathcal{A}}{2\pi\ell^{2}_{B}}
\sum_{N=1}^{\infty}
\left(
\varepsilon^{\ }_{N,-}
-
\varepsilon^{\ }_{-N}
\right)
\end{split}
\label{eq: electronic gain with Zeeman}
\end{equation}
where $\Theta$ is the Heaviside function.
The difference between the total (electronic and phonon)
energy with and without the Kekul\'e order is
\begin{equation}
\delta E=
-
2\times
\frac{\mathcal{A}}{2\pi\ell^{2}_{B}}
\Theta\left(\Delta^{\ }_{0}-\Delta^{\ }_{\mathrm{Z}}\right)
\left(
\Delta^{\ }_{0}-\Delta^{\ }_{\mathrm{Z}}
\right)
+
\frac{\mathcal{N}}{2}
\frac{K^{\ }_{\Delta}}{f^{2}_{\Delta}}
\Delta^{2}_{0}
\end{equation}
according to Eq.~(\ref{eq: elastic loss}).
Condition~(\ref{eq: condition to drop Zeeman})
follows from seeking the global minimum of 
$\delta E(\Delta^{\ }_{0})$ 
with 
$\Delta^{\ }_{0}>0$ 
and $\delta E(\Delta^{\ }_{0})<0$.

The measurements from Jiang et al.\
in Ref.~\onlinecite{Jiang07}
give the estimate
\begin{equation}
2\times\Delta^{\ }_{\mathrm{Z}}\approx
1.3\times B\hbox{ [K]}
\label{eq:estimate-Zeeman}
\end{equation}
for the Zeeman splitting.  Hence, although the
phonon-induced Kekul\'e gap~(\ref{eq:estimate-Kekule})
is larger that the one measured in Ref.~\onlinecite{Jiang07},
it is short of satisfying condition~(\ref{eq: condition to drop Zeeman}),
unless it is enhanced by another mechanism,
say by the short-range part of the Coulomb electron-electron repulsions. 
Indeed, there are contributions other than phonons that also
favor the Kekul\'e instability. For example, in
Ref.~\onlinecite{Hou07}, the nearest-neighbor (extended Hubbard)
potential was found to favor the Kekul\'e instability for strong
enough coupling. In the next section, we turn to the discussion of the
Coulomb interaction, and the issues of its symmetries and
anisotropies. We will show that the anisotropies are essential to
lower the symmetry of the problem to either a discrete symmetry or an
Abelian U(1) symmetry, so that ordering can take place in the 2D
system. The fact that a Kosterlitz-Thouless transition is seen in the
experiments of Ref.~\onlinecite{Checkelsky08} suggests that the
interactions are sufficient to boost the U(1) Kekul\'e ordering over
the Zeeman term.

\section{
Electron-electron interactions
        }
\label{sec: Electron-electron interactions}

We have ignored Coulomb electron-electron repulsions so far.
In the absence of magnetic fields and by power counting,
short-range interactions are marginally irrelevant 
while the 3D long-range Coulomb interaction is marginal.
In the latter case, it is an empirical fact that 
the Coulomb long-range interaction is marginally irrelevant
in the samples of graphene studied to this date.

The 3D long-range Coulomb interaction takes the form
\begin{equation}
\hat{H}^{\ }_{\mathrm{Cb}}=
\frac{1}{2}
\int d^{2}\boldsymbol{r}
\int d^{2}\boldsymbol{r}'
\hat{\rho}(\boldsymbol{r})
V^{\ }_{\mathrm{Cb}}(\boldsymbol{r})
\hat{\rho}(\boldsymbol{r}').
\label{eq: def long-range Cb}
\end{equation}
The local electronic density operator
\begin{equation}
\hat{\rho}(\boldsymbol{r}):=
\left(
\hat{\psi}^{\dag}
\hat{\psi}
\right)(\boldsymbol{r})
\label{eq: def long-range Cb b}
\end{equation}
is constructed from the 8-component spinor-valued operators
$\hat{\psi}^{\dag}(\boldsymbol{r})$
and
$\psi(\boldsymbol{r})$
that follow upon second-quantization of the
Dirac single-particle $8\times8$ Hamiltonian%
~(\ref{eq: def rep continuum limit}):
\begin{equation}
\hat{H}^{\ }_{0}:=
\int d^{2}\boldsymbol{r}\,
\hat{\psi}^{\dag}(\boldsymbol{r})
\mathcal{H}^{\ }_{0}(\boldsymbol{r})
\hat{\psi}(\boldsymbol{r}).
\label{eq: def hat H0}
\end{equation}
Finally, the 3D Coulomb two-body repulsive interaction is
\begin{equation}
V^{\ }_{\mathrm{Cb}}(\boldsymbol{r})=
\frac{e^{2}}{\epsilon |\boldsymbol{r}|}
\end{equation}
with $\epsilon$ the effective dielectric constant
resulting from the substrate below and the air above graphene.

The long-range Coulomb interaction~(\ref{eq: def long-range Cb})
shares the global U(8) symmetry of the coarse-grain electronic density%
~(\ref{eq: def long-range Cb b}) under
\begin{equation}
\hat{\psi}^{\dag}\to 
\hat{\psi}^{\dag}\mathcal{U}^{\dag},
\qquad
\hat{\psi}\to
\mathcal{U}
\hat{\psi},
\qquad
\mathcal{U}\in\mathrm{U(8)}.
\label{eq: U(8) sym long range Cb}
\end{equation}
This global U(8) symmetry is broken down to its 
U(4) subgroup generated by the generators in Eqs. (\ref{eq: generators U(4) sym H0-1}) and (\ref{eq: generators U(4) sym H0-2}) that leave the kinetic energy in Eq.(\ref{eq: def hat H0}) invariant.

The characteristic energy scale of the long-range 
Coulomb interaction in the presence of a magnetic field 
is estimated in Refs.~\onlinecite{Nomura06},
\onlinecite{Alicea06}, and \onlinecite{Goerbig06}
to be
\begin{equation}
120\sqrt{B}\hbox{[K]}\sim
\frac{e^{2}}{\epsilon\ell^{\ }_{B}}<
\hbar\omega^{\ }_{\mathrm{c}}\sim
400\sqrt{B}\hbox{[K]}
\label{eq: estimate LR CB}
\end{equation}
in units of Kelvin
(with $B$ measured in units of the Tesla).
This makes the long-range
Coulomb interaction the second largest energy scale
for the range of magnetic fields relevant to 
Ref.~\onlinecite{Checkelsky09}
after that arising from 
the kinetic energy~(\ref{eq: def hat H0}).

The U(8) symmetry~(\ref{eq: U(8) sym long range Cb})
of the long-range Coulomb interaction is not exact. 
On the length scale of the honeycomb lattice spacing,
the local electronic density operator is not given by the coarse-grain
approximation~(\ref{eq: def long-range Cb b})
but it resolves the sublattice-1/2 and valley-1/2 quantum numbers
down to the point group symmetry of the tight-binding model.
Consequently, the on-site or the nearest-neighbor  
density-density repulsive interactions break the U(8) symmetry
down to the product between the
U(2) subgroup that generates the conservation of the
charge and of the spin-1/2 quantum numbers on the one hand
and of the discrete point group on the other hand. 
The spin-1/2 SU(2) symmetry is in turn broken by the
Zeeman interaction~(\ref{eq: def Zeeman interaction}).

The characteristic energy scale for the Coulomb repulsion
on the scale of the lattice spacing must be of the order
\begin{equation}
\frac{e^{2}}{\ell^{\ }_{B}}\times f(\mathfrak{a}/\ell^{\ }_{B})
\end{equation} 
with $f$ a dimensionless function obeying $f(x\to0)=0$.
If the asymptotic expansion $f(x\to0)\sim x$ holds, 
it then follows that the characteristic energy scale
of the lattice Coulomb interaction grows linearly with $B$
\begin{equation}
\frac{e^{2}}{\ell^{\ }_{B}}\times f(\mathfrak{a}/\ell^{\ }_{B})\sim
B
\end{equation} 
and could thus be comparable to the Zeeman splitting~(\ref{eq:estimate-Zeeman})
or to the phonon-induced Kekul\'e gap~(\ref{eq:estimate-Kekule}).
We shall assume this to be the case.

Because of the Mermin-Wagner theorem,
the long-range Coulomb interaction
cannot break spontaneously the combined U(4) spin and valley symmetry
of the kinetic energy $\mathcal{H}^{\ }_{0}$
from Eq.~(\ref{eq: def hat H0})
at any non-vanishing temperature. 
However, any subdominant interaction
that breaks the U(4) symmetry down to
a U(1) or a finite subgroup of U(4)
(say its center $\mathbb{Z}^{\ }_{4}$)
can establish a non-vanishing-temperature transition
to a quasi-long-range or long-range-order phase,
respectively. Hence, any of the Zeeman, 
the phonon-electron, or the short-range Coulomb interactions
can induce a non-vanishing-temperature phase transition. 
This brings about two questions. What is the leading
instability and at what temperature does the leading
instability takes place? 

First, since any finite-temperature instability 
must be associated to a discrete symmetry
or to a U(1) symmetry subgroup of U(4), the empirical observation
of a KT-like transition rules out any discrete symmetry subgroup.
This empirical fact can be understood at the mean-field
level from three observations.
(i) We showed in Secs.~\ref{sec: The Kekule instability} 
and \ref{sec: Competing Zeeman energy} 
that phonons favor a Kekul\'e instability
with a gap larger (but not substantially larger) 
than the Zeeman splitting.
(ii) Moreover, a repulsive nearest-neighbor Coulomb interaction reinforces
the phonon-induced instability.~\cite{Hou07} 
(iii) An exhaustive list of instabilities in graphene can be found in 
Ref.~\onlinecite{Ryu09}.
Because the cyclotron energy is the dominant energy scale for the
relevant range of magnetic field, we can ignore U(1) instabilities
associated to superconductivity. This still leaves several
competing U(1) order parameters (see Table 1 in Ref.~\onlinecite{Ryu09})
but of these only one, the Kekul\'e instability, 
is also favored by phonons.

Second, we have seen in Eq.~(\ref{eq:Landau-spectrum with Zeeman}) 
that the Zeeman interaction $\mathcal{H}^{\ }_{\mathrm{Z}}$
opens a single-particle gap at half-filling in the Landau level
$\nu=0$,
i.e., the Slater determinant made of all 
$\nu=0$ single-particle energy eigenstates
of $\mathcal{H}^{\ }_{0}$
in Eq.~(\ref{eq: def rep continuum limit})
with negative eigenenergies 
has a larger energy than that of
$\mathcal{H}^{\ }_{0}+\mathcal{H}^{\ }_{\mathrm{Z}}$
at half-filling. The Zeeman term is also a symmetry breaking field
that lowers the U(4) symmetry of 
$\mathcal{H}^{\ }_{0}$
in Eq.~(\ref{eq: def rep continuum limit}).
Any of the 15 traceless generators of the U(4)
symmetry~(\ref{eq: generators U(4) sym H0})
would do the same, or, more generally 
any 15-component uniform real-valued
``SU(4)-symmetry-breaking magnetic field''
$b=(b^{\ }_{\mathrm{a}})$ in
\begin{subequations}
\begin{eqnarray}
&&
\mathcal{H}(b):=
\sum_{\mathrm{a}=1}^{15}
b^{\ }_{\mathrm{a}} T^{\ }_{\mathrm{a}},
\\
&&
T^{\ }_{1}:=
\frac{1}{\sqrt{16}} 
X^{\ }_{001},
\cdots,
T^{\ }_{15}:=
\frac{1}{\sqrt{16}} 
X^{\ }_{303},
\end{eqnarray}
\end{subequations}
opens a single-particle gap if added to $\mathcal{H}^{\ }_{0}$.
[The factor of $1/\sqrt{16}$ insures the normalization
$\mathrm{tr}\, T^{\ }_{\mathrm{a}}T^{\ }_{\mathrm{b}}=
\delta^{\ }_{\mathrm{a}\mathrm{b}}/2$
for the 15 generators $T^{\ }_{\mathrm{a}}$ of SU(4).]

These single-particle considerations are also valid
in the presence of the Coulomb interaction.
As is the case in double-layer non-relativistic
quantum Hall systems,%
~\cite{Wen93,Moon95,Yang96}
the long-range Coulomb interaction in the
$\nu=0$ Landau level is minimized by 
breaking spontaneously the SU(4) symmetry in
Eq.~(\ref{eq: generators U(4) sym H0})
in the same way as a quantum SU(4) ferromagnet would do,%
~\cite{Nomura06,Alicea06,Goerbig06}
i.e., in such a way that
$\hat{\psi}^{\dag}(\boldsymbol{r})
 \mathcal{H}(b)
 \hat{\psi}(\boldsymbol{r})$
acquires spontaneously a finite expectation value 
independent of $\boldsymbol{r}$
in the many-body ground state for some 
``SU(4)-symmetry-breaking magnetic field'' 
$b=(b^{\ }_{\mathrm{a}})$.
Such a many-body ground state is symmetric
under any exchange of a pair of its SU(4) indices. Consequently, the Pauli principle can maximize the
spatial separation of electrons to lower
the Coulomb interaction without penalizing the kinetic energy
due to the magnetic field.

If the Coulomb interaction is treated in a mean-field approximation
by which $\hat{H}^{\ }_{\mathrm{Cb}}\to\mathcal{H}(b)$
for some ``SU(4)-symmetry-breaking magnetic field'' 
$b=(b^{\ }_{\mathrm{a}})$,
the presence of a single-particle mean-field gap justifies integrating 
the fermions in a gradient expansion.
If this is done at any small but non-vanishing temperature,
we can ignore quantum fluctuations to a first approximation.
There follows an effective classical theory that describes how  
small thermal fluctuations about the SU(4)-symmetry-broken
ground state restore the full SU(4) symmetry,
as dictated by the Mermin-Wagner theorem 
(we always assume that the dominant cyclotron energy prevents
any superconducting instability). On symmetry grounds, this
classical theory is a non-linear-sigma model (NLSM) 
on the target manifold $\mathrm{G/H}$ given by
the $N-1$-dimensional complex projective space
\begin{eqnarray}
\mathbb{C}\mathrm{P}^{N-1}&\simeq&
\hbox{U($N$)/U($N-1$)$\times$U(1)}
\nonumber\\
&\simeq&
\hbox{SU($N$)/S\big(U($N-1$)$\times$U(1)\big)}
\end{eqnarray}
with $N=4$
(see Refs.~\onlinecite{Eichenherr78}, \onlinecite{Dadda78},
\onlinecite{Golo78}, and \onlinecite{Polyakov87} 
for the introduction and overview
of $\mathbb{C}\mathrm{P}^{N-1}$ NLSM).
The little group H is here the largest subgroup of G that
leaves the ground state 
$\psi(b)$ of $\mathcal{H}(b)$
invariant. It is made of the
direct product of the subgroup SU(3) 
spanned by all unitary transformation
in the subspace orthogonal to the direction of 
$\psi(b)$ in the combined spin and valley subspace
and the subgroup U(1) spanned by all rotations about 
$\psi(b)$ in the very same spin and valley subspace.
If we also include the 
symmetry-breaking subleading interactions
(the Zeeman, electron-phonon, and electron-electron
short-range Coulomb interactions),
this effective theory depends on two
dimensionless couplings. 
There is the reduced temperature $t$.
There also is the coupling $\lambda\ll t$
that breaks the SU(4)
symmetry to subleading order 
so as to enforce the desired pattern%
~(\ref{eq: desired pattern symmetry breaking})
of explicit symmetry breaking.
By power counting, $t$ is marginal while $\lambda$ is strongly relevant.
(These couplings have already been renormalized by quantum fluctuations.)

On the one hand, if we ignore the anisotropy $\lambda$, 
the reduced temperature $t$ flows away from 
the unstable infra-read fixed point $t=0$
to strong coupling with the beta function 
(see Refs.~\onlinecite{Brezin80},
\onlinecite{Hikami81},
\onlinecite{Hikami83},
and
\onlinecite{Wegner89})
\begin{equation}
\beta^{\ }_{t}(t)\equiv
\mathfrak{a}\frac{\partial t}{\partial\mathfrak{a}}=
\sum_{n=1,2,3,\cdots}^{\infty}
\beta^{\ }_{n} t^{n}
\end{equation}
that vanishes to order $n=1$ and is given by
the quadratic Casimir in the adjoint representation
$\mathrm{C}^{\ }_{v}$
\begin{equation}
\beta^{\ }_{2}=
k\times\mathrm{C}^{\ }_{v}=
k\times N
\end{equation}
with $N=4$ to one loop order ($n=2$), 
up to the multiplicative factor $k$ that depends
on the convention made for the normalization of the 
reduced temperature
($k=1$ in Refs.~\onlinecite{Brezin80},
\onlinecite{Hikami81},
\onlinecite{Hikami83},
and
\onlinecite{Wegner89}).
Correspondingly, as $t$ grows under this renormalization flow
the correlation length $\xi(t)$
decreases,
\begin{equation}
\xi(t)\sim
\mathfrak{a}
\exp
\left(
+
\frac{1/(k\times N)}{t}
\right)
\end{equation}
with $N=4$. On the other hand, the mass scale $\lambda^{1/2}$ grows
away from the Gaussian (unstable) fixed point $t=\lambda=0$
to leading order in the flow of the renormalization group
according to 
\begin{equation}
\beta^{\ }_{\lambda}(\lambda)\equiv
\mathfrak{a}\frac{\partial \lambda}{\partial\mathfrak{a}}=
2\lambda.
\end{equation} 
The symmetry crossover is estimated from the condition
$(\xi/\mathfrak{a})^{-2}\approx\lambda\ll1$, i.e.,%
~\cite{Pelcovits76,Khokhlachev76,Hikami80}
\begin{equation}
t\approx
\frac{2/(k\times N)}{\ln(c/\lambda)}\ll1
\label{eq: condition for sym cross}
\end{equation}
with $\lambda\ll c$ a positive constant of order one
fixed by the short-distance physics.
Once Eq.~(\ref{eq: condition for sym cross}) 
is fulfilled, it is the classical 
renormalization group flow in the
unbroken subgroup of U(1) of SU(4) that takes over.

We conclude that the Kekul\'e mechanism can explain
the experiment of Ref.~\onlinecite{Checkelsky09}
if 
(i) the phonon-induced Kekul\'e instability can
be boosted by the nearest-neighbor
short-range Coulomb interaction
that favors the Kekul\'e instability
so as to satisfy Eq.~(\ref{eq: condition to drop Zeeman}),
(ii) the renormalized SU(4)-symmetric energy scale
$J^{\ }_{\mathrm{SU(4)}}$ 
[as opposed to the bare estimate~(\ref{eq: estimate LR CB})]
that enters the reduced temperature in the NLSM satisfies
\begin{eqnarray}
T^{\ }_{c}&\approx&
\frac{1/(k\times4)}{\ln 50}\times J^{\ }_{\mathrm{SU(4)}}\,[K]
\nonumber\\
&\approx&
k^{-1}\times 0.12\times J^{\ }_{\mathrm{SU(4)}}\,[K]
\end{eqnarray}
since $\lambda$ is of order 
$\mathfrak{a}/\ell^{\ }_{B}\sim1/50$ for the relevant range
of magnetic fields and if $c=1$ 
in Eq.~(\ref{eq: condition for sym cross}). 
This estimate will be lowered by quantum fluctuations and by
disorder (as described in Sec.~\ref{sec: The Kekule instability}
for the disorder).

\section{
Discussion
        }
\label{sec: Discussion} 

In summary, we have argued that sufficiently large magnetic fields and
sufficiently small temperatures (inside the shaded region of the phase
diagram of Fig.~\ref{fig:2Dphase-deiagram}) stabilize the Kekul\'e
quasi-long-range order in graphene.  This quasi-long-range order can
be destroyed by the unbinding of vortices due to either thermal- or
disorder-induced fluctuations.  The role of the magnetic field $B$ is
here 2-fold.  First, $B$ enhances the density of states at the Dirac
point, thereby favoring the Kekul\'e distortion with the help of
phonons. Second, a large $B$ prevents the disorder from filling a
single-particle gap by suppressing the Landau degeneracy.  The
transition belongs to the 2D classical random phase $XY$ model and is
characterized by the deconfinement vortices for sufficiently small
magnetic fields or sufficiently large temperatures (outside the shaded
region of the phase diagram of Fig.~\ref{fig:2Dphase-deiagram}). Each
of these vortices, as we have shown here, bind a fraction of the
electron charge for any magnetic field strength.

\section*{
Acknowledgments } 

This work was supported in part by DOE Grant DEFG02-06ER46316
(C.-Y.~H. and C.~C.). C.~M.\ acknowledges the kind hospitality of the
Condensed Matter Theory Laboratory in RIKEN and the Condensed Matter
Theory Visitor's Program at Boston University. We thank Y. Kopelevich,
N. P. Ong, J. G. Checkelsky, S. Ryu, D.-H. Lee, K. Nomura, and
A. Furusaki for useful and stimulating discussions.

\appendix

\section{ Stiffness }
\label{sec:Stiffness}

In this appendix, we start from
Hamiltonian~(\ref{eq:def-mathcal{H}}) 
with the uniform magnetic field 
$
B=
\partial A^{\ }_{y}/\partial^{\ }_{x}
-
\partial A^{\ }_{x}/\partial^{\ }_{y}
$
and with the uniform Kekul\'e order parameter
$\Delta=\Delta^{\ }_{0}e^{i\theta^{\ }_{0}}$.
We choose to work in the 
Landau gauge $\boldsymbol{A}=(-By,0)$.
We are after the Kekul\'e phase stiffness.

To this end, we first need the eigenfunctions and
eigenenergies of Hamiltonian~(\ref{eq:def-mathcal{H}}).
That the eigenenergies are given by
Eq.~(\ref{eq:Landau-spectrum})
follows from the fact that the uniform Kekul\'e order 
parameter anticommutes with the covariant derivative.
Hence, any positive eigenenergy of 
$\mathcal{H}^{2}$ is nothing but the sum of
two positive terms adding in quadrature.
One is the eigenvalue of the square of the covariant derivatives 
in Eq.~(\ref{eq:def-mathcal{H}}).
The other is the eigenvalue of the square of the uniform 
Kekul\'e order parameter
in Eq.~(\ref{eq:def-mathcal{H}}).
To obtain the (degenerate) eigenfunctions of any Landau level%
~(\ref{eq:Landau-spectrum}),
we anticipate that any eigenfunction of Eq.~(\ref{eq:def-mathcal{H}})
factorizes into the $x$-dependent phase
$e^{{i}kx}$ times a 4-component spinor
that depends solely on $y$ through the dimensionless co-ordinate
$\xi=(y/\ell^{\ }_{B})+k\ell^{\ }_{B}$.
As it does for the conventional (non-relativistics) Landau levels,
the wave number $k$ encodes a degeneracy of the spectrum%
~(\ref{eq:Landau-spectrum})
that scales with the area of the system.
An additional (relativistic) source of degeneracy of the spectrum%
~(\ref{eq:Landau-spectrum})
arises because of the dimensionality 4 of the Dirac matrices.
This finite degeneracy in the sole presence of a uniform magnetic field
is selectively lifted by the uniform Kekul\'e order parameter. 
Indeed, the single-particle eigenstates with $N=0$ 
are non-degenerate for each wave number $k$ 
as a result of the Kekul\'e instability and given by
\begin{equation}
\Psi^{\ }_{k,0,\pm}(x,\xi)=
\frac{e^{{i}kx}}{\sqrt{2}}
\begin{pmatrix}
\varphi^{\ }_{0}(\xi)
\\
0
\\
\pm 
\varphi^{\ }_{0}(\xi) 
e^{-{i}\theta^{\ }_{0}}
\\
0
\end{pmatrix}.
\end{equation}
On the other hand,
for any wave number $k$, positive integer $N=1,2,\cdots$,
and sign $\pm$, the pair of orthonormal eigenfunctions
\begin{equation}
\begin{split}
\Psi^{\ }_{k,N,\pm}(x,\xi)=&\,
\frac{e^{{i}kx}}{\sqrt{2}}
\begin{pmatrix}
\frac{
+\hbar\omega^{\ }_{\mathrm{c}}\sqrt{N}
     }
     {
\varepsilon^{\ }_{N,+}
     }  
\varphi^{\ }_{N}(\xi)
\\
\pm\varphi^{\ }_{N-1}(\xi)
\\
0
\\
\frac{
\Delta^{\ }_{0}
     }
     {
\varepsilon^{\ }_{N,+}
     }  
\varphi^{\ }_{N-1}(\xi) 
e^{-i \theta^{\ }_{0}}
\end{pmatrix}
\end{split}
\end{equation}
and
\begin{equation}
\begin{split}
\tilde{\Psi}^{\ }_{k,N,\pm}(x,\xi)=&\,
\frac{e^{{i}kx}}{\sqrt{2}}
\begin{pmatrix}
\frac{
\Delta^{\ }_{0}}
     {
\varepsilon^{\ }_{N,+}
     }  
\varphi^{\ }_{N}(\xi)
\\
0
\\
\pm\varphi^{\ }_{N}(\xi)  
e^{-{i}\theta^{\ }_{0}}
\\
\frac{
-\hbar\omega^{\ }_{\mathrm{c}}\sqrt{N}
     }
     {
\varepsilon^{\ }_{N,+}
     }
\varphi^{\ }_{N-1}(\xi) 
e^{-{i}\theta^{\ }_{0}}
\end{pmatrix}
\end{split}
\end{equation}
remain 2-fold degenerate in spite of the Kekul\'e instability. Here,
$\varphi^{\ }_{N}$ with $N=0,1,2,\dots$ 
are the orthonormal eigenfunctions of the
one-dimensional quantum harmonic oscillator.

Next, the Kekul\'e phase stiffness can be computed by adding all the changes
in the negative single-particle energy
levels~(\ref{eq:Landau-spectrum}) up to the Fermi level $\mu=0$ that
are induced by the twist
$\Delta\to\Delta\exp({i}\boldsymbol{q}\cdot\boldsymbol{r})$ in the
Kekul\'e order parameter. 

Before doing this, it is convenient to gauge out
the spatial dependence of the Kekul\'e order parameter 
with the help of the pure axial gauge transformation 
${\Psi'}\equiv U\Psi$, where 
$U\equiv e^{-{i}\boldsymbol{q}\cdot\boldsymbol{r}\gamma^{\ }_{5}/2}$.
Under this transformation, the Hamiltonian becomes
$\mathcal{H}'= \mathcal{H}+\mathcal{V}$, where
$\mathcal{H}$ is given by~(\ref{eq:def-mathcal{H}}) 
with the Landau spectrum~(\ref{eq:Landau-spectrum}) and
$\mathcal{V}=  v^{\ }_{\mathrm{F}}\,
 \boldsymbol{q}\cdot\boldsymbol{\alpha}\,
 \gamma^{\ }_{5}/2$.

We treat $\mathcal{V}$ as a perturbation, 
assume that 
$\hbar v^{\ }_{\mathrm{F}}|\boldsymbol{q}|$ 
is small compared to $\Delta^{\ }_{0}$,
and compute the change in the spectrum of $\mathcal{H}$
induced by $\mathcal{V}$
up to second order in degenerate perturbation theory. 
Because the perturbation $\mathcal{V}$ does not
couple eigenstates with different wave number $k$, the degeneracy
of each unperturbed Landau level~(\ref{eq:Landau-spectrum})
is conserved. Furthermore, 
these second-order shifts $\varepsilon^{(2)}_{k,N,\pm}$
cancel pairwise for all 2-fold degenerate levels:
$\varepsilon^{(2)}_{k,N,+}+\varepsilon^{(2)}_{k,N,-}=0$ 
when $N=1,2,\cdots$. 
Hence at half-filling ($\mu=0$), the second order shifts from all states with
$N=1,2,\cdots$ cancel pairwise, leaving the unpaired $N=0$ states as
sole contributors to the total energy shift:
\begin{equation}
\delta E=
2\times \sum_k \varepsilon^{(2)}_{k,0,-}
\end{equation}
where the factor of 2 accounts for the spin-1/2 degeneracy, and the
second-order correction, 
\begin{equation}
\varepsilon^{(2)}_{k,0,-}=
v^{2}_{\mathrm{F}}|\boldsymbol{q}|^{2}   
\frac{ \Delta^{\ }_{0}}{2 \omega^{2}_{c}}
=\frac{\ell^2_B }{4 }
|\boldsymbol{q}|^{2}   
\Delta^{\ }_{0},
\end{equation}
are independent of $k$. The total energy shift per unit
area $\delta E/\mathcal{A}$ can be obtained by accounting for the density
of states per area and per spin $1/(2\pi\ell^{2}_{B})$ at the
unperturbed energy $-\Delta^{\ }_{0}$, i.e.,
\begin{equation}
\frac{\delta E}{\mathcal{A}}=
\frac{\Delta^{\ }_{0}}{4\pi}|\boldsymbol{q}|^{2}.
\label{eq:deltaE-per-Area}
\end{equation}
If we define the stiffness by
\begin{equation}
\frac{\delta E}{\mathcal{A}}\equiv
\frac{J}{2}|\boldsymbol{q}|^{2},
\label{eq: def J}
\end{equation}
we deduce that
\begin{equation}
J=
\frac{\Delta^{\ }_{0}}{2 \pi}.
\label{eq:estimate-for-J-appendix} 
\end{equation}
Because the self-consistent Kekul\'e gap $\Delta^{\ }_{0}$ induced by
a magnetic field and an electron-phonon interaction scales linearly in
$B$, the spin-stiffness also scales linearly with $B$.

\section{ Zero modes }
\label{sec:zero-mode}

\begin{widetext}

The zero modes of the Dirac Hamiltonian~(\ref{eq:def-mathcal{H}}), 
when the Kekul\'e order parameter $\Delta$
is defective in that it carries a vortex of vorticity $\pm1$, 
can be found analytically and shown to exist for all strengths of the
applied magnetic field. 
In this appendix, we choose to work in the symmetric gauge
$\boldsymbol{A}=\frac{B}{2}\;(-y,x)$, which we write as the complex
number $A={i}Bz/2$ with $z\equiv x+{i}y$,
and seek the solutions (zero modes) to
\begin{equation}
\begin{pmatrix}
0 
& 
\hbar v^{\ }_{\mathrm{F}}
\left(
-
2i\partial^{\ }_{z}
+
\frac{i\bar{z}\ell^{2}_{B}}{2}
\right)
& 
\Delta
& 
0
\\
\hbar v^{\ }_{\mathrm{F}}
\left(
-
2i\partial^{\ }_{\bar z}
-
\frac{{i}z\ell^{2}_{B}}{2}
\right)
&
0
& 
0 
&
\Delta\\
\bar \Delta 
& 
0 
& 
0 
& 
\hbar v^{\ }_{\mathrm{F}}
\left(
+2i\partial^{\ }_{z}
-
\frac{i\bar{z}\ell^{2}_{B}}{2}
\right)
\\
0 
& 
\bar{\Delta} 
&
\hbar v^{\ }_{\mathrm{F}}
\left(
+
2i\partial^{\ }_{\bar z}
+
\frac{{i}z\ell^{2}_{B}}{2}
\right)
&
0
\end{pmatrix}
\Psi=0.
\label{eq: def zero mode}
\end{equation}
The spinor 
\begin{subequations}
\begin{equation}
\Psi^{\ }_{\mathrm{A}}(\boldsymbol{r})=
\begin{pmatrix}
0
\\
u(\boldsymbol{r})
\\
v(\boldsymbol{r})
\\
0
\end{pmatrix}
\end{equation}
is a zero mode supported on sublattice $\mathrm{A}$ if
\begin{equation}
\begin{split}
&
\hbar v^{\ }_{\mathrm{F}}
\left(
-
2i\partial^{\ }_{z}
+
\frac{i\bar z\ell^{2}_{B}}{2}
\right)u
+
\Delta 
v=
0,
\\
&
\bar{\Delta} u 
+
\hbar v^{\ }_{\mathrm{F}}
\left(
2i\partial^{\ }_{\bar{z}}
+
\frac{{i}z\ell^{2}_{B}}{2}
\right)v
=
0,
\end{split}
\end{equation}
\end{subequations}
and $u$ and $v$ are normalizable.
Switching to polar coordinates and considering an anti-vortex
(vorticity $n^{\ }_{\mathrm{v}}=-1$) 
for concreteness, i.e., 
$\Delta= 
 \Delta(\rho,\theta)=
 \Delta^{\ }_{0}\,e^{{i}\theta^{\ }_{0}}\, e^{-{i}\theta}$, 
the conditions on the components $u$ and $v$ become
\begin{subequations}
\label{eq: eqs for -1 antivortex}
\begin{eqnarray}
&&
-i\hbar v^{\ }_{\mathrm{F}}
\left(
\partial^{\ }_\rho 
-
\frac{i}{\rho}\partial^{\ }_{\theta}
- 
\frac{\rho}{2\ell^{2}_{B}}
\right)u
+
\Delta^{\ }_{0}\,e^{{i}\theta^{\ }_{0}} v=
0,
\label{eq:u-v-1}
\\
&&
\Delta^{\ }_{0}\,e^{{-i}\theta^{\ }_{0}} u 
+
i\hbar v^{\ }_{\mathrm{F}}
\left(
\partial^{\ }_\rho 
+
\frac{i}{\rho}\partial^{\ }_{\theta}
+ 
\frac{\rho}{2\ell^{2}_{B}}
\right)v=
0.
\label{eq:u-v-2}
\end{eqnarray}
\end{subequations}
The homogeneous system~(\ref{eq: eqs for -1 antivortex})
of first-order partial differential equations
admits $\theta$-independent solutions $u(\rho)$ and
$v(\rho)$. Indeed,
solving for $u$ in terms of $v$ in Eq.~(\ref{eq:u-v-2}) gives
\begin{subequations}
\begin{equation}
u=
-i\hbar v^{\ }_{\mathrm{F}}\Delta^{-1}_{0}\,e^{{i}\theta^{\ }_{0}}
\left(
\partial^{\ }_\rho 
+ 
\frac{\rho}{2\ell^{2}_{B}}
\right)v,
\end{equation} 
and, after substitution into Eq.~(\ref{eq:u-v-1}),
\begin{eqnarray}
-
\left(\ell^{\ }_{B}\,\partial^{\ }_{\rho}\right)^2v 
+ 
\frac{1}{4}\left(\frac{\rho}{\ell^{\ }_{B}}\right)^2\,v
=-
\left(
\frac{1}{2}
+
p
\right)\,v,
\label{eq:v-only}
\end{eqnarray}
where
\begin{equation}
p=
\left(\frac{\ell_B \Delta_0}{\hbar v^{\ }_{\mathrm{F}}}\right)^2
=
\frac{
2\Delta^{2}_{0}
     }
     {
(\hbar\omega^{\ }_{\mathrm{c}})^{2}
     }.
\end{equation}
\end{subequations}

Equation~(\ref{eq:v-only}) would be identical to the second-order
differential equation describing a 1D quantum harmonic oscillator were
it not for the fact that the radial coordinate $\rho$ is always
positive.  Equation~(\ref{eq:v-only}) thus admits normalizable
solutions unavailable to the 1D quantum harmonic oscillator, i.e., any
normalizable solution on the half-line that blows up when
$\rho\to-\infty$ is here allowed.  The solutions of
Eq.~(\ref{eq:v-only}) for $\rho>0$ are parabolic cylinder functions
$D^{\ }_{-p}(\rho/\ell_B)$,~\cite{Gradshteyn00} parametrized by the
dimensionless ratio $p$.  We conclude, when the Kekul\'e order
parameter admits an anti-vortex of charge $n^{\ }_{\mathrm{v}}=-1$,
that a normalizable zero mode of Eq.~(\ref{eq: def zero mode}) can be
expressed in terms $u$ and $v$ satisfying Eqs.~(\ref{eq:u-and-v1}) and
(\ref{eq:u-and-v2}). The solution for the case of vorticity $n^{\
}_{\mathrm{v}}=+1$ is analogous, but with a zero mode supported on
sublattice $\mathrm{B}$.

\end{widetext}

\end{document}